\documentclass[prd,twocolumn,showpacs,preprintnumbers,superscriptaddress,ams
math]{revtex4}
\begin{document}
\title{On Magnetic solution to 2+1 Einstein--Maxwell
gravity}
\author{Mauricio Cataldo}
\altaffiliation{mcataldo@ubiobio.cl}%0915
\affiliation{Departamento de F\'\i sica, Facultad de Ciencias,
Universidad del B\'\i o--B\'\i o, Avenida Collao 1202, Casilla
5-C, Concepci\'on, Chile.\\}
\author{Juan Cris\'ostomo}
\altaffiliation{juan.crisostomo@ucv.cl} \affiliation{Instituto de
F\'\i sica, Facultad de Ciencias B\'asicas y Matem\'aticas,
Pontificia Universidad Cat\'olica de Valpara\'\i so, Avenida
Brasil 2950, Valpara\'\i so, Chile.}
\author{Sergio del Campo}
\altaffiliation{sdelcamp@ucv.cl} \affiliation{Instituto de F\'\i
sica, Facultad de Ciencias B\'asicas y Matem\'aticas, Pontificia
Universidad Cat\'olica de Valpara\'\i so, Avenida Brasil 2950,
Valpara\'\i so, Chile.}
\author{Patricio Salgado}
\altaffiliation{pasalgad@udec.cl}%0915
\affiliation{Departamento de F\'\i sica, Facultad de Ciencias
F\'\i sicas y Matem\'aticas, Universidad de Concepci\'on, Casilla
160-C, Concepci\'on, Chile.}
\date{\today}
\begin{abstract}
The three--dimensional magnetic solution to the Einstein--Maxwell
field equations have been considered by some authors. Several
interpretations have been formulated for this magnetic spacetime.
Up to now this solution has been considered as a two--parameter
self--consistent field. We point out that the parameter related to
the mass of this solution is just a pure gauge and can be rescaled
to minus one. This implies that the magnetic metric has really a
simple form and it is effectively one-parameter solution, which
describes a distribution of a radial magnetic field in a 2+1
anti--de Sitter background space--time. We consider an alternative
interpretation to the Dias--Lemos one for the magnetic field
source.

\end{abstract}
\maketitle The 2+1 magnetic solution to the Einstein--Maxwell
field equations has been studied by some authors. The static
solution has been found by Clement~\cite{Clement},
Peld\'{a}n~\cite{Peldan}, Hirschmann and Welch~\cite{H-W} and
Cataldo and Salgado~\cite{Cataldo}, using different procedures.
The generalization to the rotating case was done by Dias and Lemos
\cite{Lemos}. This solution may be written in the form
\begin{eqnarray}\label{H-W}
ds^2=- \left(\frac{r^2}{l^2}-M \right)d t^2 +
\nonumber \\
\frac{r^2 d r^2}{\left(\frac{r^2}{l^2}- M \right)\left( r^2+Q^2_m
\ln \left| \frac{r^2}{l^2}-M \right| \right)} + \nonumber
\\
\left( r^2+Q^2_m \ln \left| \frac{r^2}{l^2}-M \right| \right) d
\phi^2,
\end{eqnarray}
where $l$ is the radius of a pseudo--sphere related to the
cosmological constant via $l=-1/\sqrt{\Lambda}$, $Q_{m}$ and $M$
are self--consistent integration constants of the
Einstein--Maxwell field equations. The vector potential $1$--form
of this gravitational field is given by
\begin{eqnarray*}
A=\frac{Q_m}{2} \, \ln \left|\frac{r^2}{l^2}-M \right| \, d \phi.
\end{eqnarray*}
When $Q_m=0$ the metric~(\ref{H-W}) reduces to the nonrotating
three--dimensional Ba\~nados--Teitelboim--Zanelli black
hole~\cite{Teitelboim1}, where $M$ is the mass of this uncharged
metric, which has an event horizon at $r=\sqrt{M} l$. Let us study
the behavior of this Einstein--Maxwell field. We shall consider
the values of the $r$--coordinate for which the component $g_{\phi
\phi}$ becomes zero. This occurs to be for some value of
$r=\overline{r}$, which satisfies the constraint
\begin{eqnarray}\label{constraint}
\overline{r}^{\,2} + Q^2_m \ln \left|
\frac{\overline{r}^{\,2}}{l^2}-M \right|=0.
\end{eqnarray}
This equation implies that $\overline{r}$ is constrained to be
between
\begin{eqnarray}\label{RightInterval}
l \, \sqrt{M}<\overline{r} \leq l \sqrt{M+1}.
\end{eqnarray}
The metric~(\ref{H-W}) appears to change the signature at
$r=\overline{r}$. This indicates us that we are using an incorrect
extension. The correct one can be found setting
\begin{eqnarray}\label{cambiocoordenado}
x^2=r^2-\overline{r}^{\, 2},
\end{eqnarray}
since the physical space--time has sense only for $r \geq
\overline{r}$ and we have $0 \leq x < \infty$. Taking into account
the constraint~(\ref{constraint}), the metric~(\ref{H-W}) becomes
\begin{equation}\label{metricatransformada}
ds^{2}=-\left(\frac{x^{2}}{l^{2}}+\frac{\alpha^{2}}{l^{2}}\right)dt^{2}+
\frac{l^{2}x^{2}dx^{2}}{(x^{2}+\alpha^{2})F^{2}(x)}+F^{2}(x)
d\phi^{2},
\end{equation}
where $\alpha^{2}=\overline{r}^{2}-l^{2}M$ and the function
$F^{2}(x)$ is defined as
\begin{equation}
F^{2}(x)=x^{2}+Q^{2}_{m}\ln\left(1+\frac{x^{2}}{\alpha^{2}}\right)
\end{equation}
This metric is horizonless, without curvature singularities and in
particular, there is no a magnetically charged
three--dimensionally black hole~\cite{H-W}. The presented magnetic
solution shows a strange behavior. As the parameter $Q_{m}$,
related to the strength of the magnetic field, goes to zero we
should recover the Ba\~nados--Teitelboim--Zanelli black hole, but
it does not occur. Since, in this case ``the limit of a theory is
not the theory of the limit". Surprisingly, this strange behavior
can be eliminated by introducing a new set of coordinates.
Effectively, making the following rescaling transformations
\begin{eqnarray}\label{15C-S-H-Wtransformation}
&& t^{\prime 2}=\frac{\overline{r}^{\, 2}-Ml^2}{l^2} \, t^2,
\,\,\,\,\, r^{\prime 2}=\frac{l^2}{\overline{r}^{\, 2}-Ml^2} \,
x^2, \nonumber \\ &&\phi^{\prime 2}=\frac{\overline{r}^{\,
2}-Ml^2}{l^2} \, \phi^2,
\end{eqnarray}
and introducing them into Eq.~(\ref{metricatransformada}), we
obtain the following metric
\begin{equation}\label{SolucionFisica}
ds^{2}=-\left(\frac{r^{\prime\, 2}}{l^{2}}+1 \right)dt^{\prime\,
2}+\frac{r^{\prime\, 2}dr^{\prime\, 2}}{(\frac{r^{\prime\,
2}}{l^{2}}+1)F^{\prime\, 2}(r^{\prime\,})}+F^{\prime\,
2}(r^{\prime\,})d\phi^{\prime\, 2},
\end{equation}
where
\begin{equation}
F^{\prime\, 2}(r^{\prime\,})=r^{\prime\,
2}+\tilde{q}^{2}_{m}\ln\left(\frac{r^{\prime\, 2}}{l^{2}}+1
\right),
\end{equation}
and $\tilde{q}^2_m= Q^2_m e^{\overline{r}^{\,2}/Q^2_m}$. In the
present form, the constant $M$ has been eliminated and the
metric~(\ref{SolucionFisica}) has one integration constant $q_m$.
This parameter is well behaved for $\overline{r}=0$ and the
magnetic field can be switched off without any problem. When
$\tilde{q}_m=0$, the anti--de Sitter space is obtained. Clearly
the metric~(\ref{SolucionFisica}) is a particular solution of the
line element~({\ref{H-W}), where we need to put $M=-1$. This
implies that the parameter related to the mass of this solution is
just a pure gauge and it can be rescaled to the value $- 1$. This
agrees with the Dias--Lemos result~\cite{Lemos}, who have shown
that the mass of the magnetic solution~(\ref{metricatransformada})
is negative. However, the examined by authors three--dimensional
static magnetic field is still a two--parameter solution, since
the mass is considered a free parameter (see their Eq. (3.24) with
$\Omega=J=Q_e=0$). Thus, the metric~(\ref{metricatransformada}) is
really a one--parameter solution with a distribution of a radial
magnetic field in a 2+1 anti--de Sitter background, which takes
the form of Eq.~(\ref{SolucionFisica}). This metric can be
considered as the general ``physical solution" to the
self--consistent problem for a superposition of a radial magnetic
field and a 2+1 Einstein static gravitational field. Clearly the
metric~(\ref{SolucionFisica}) is not a magnetically charged
three--dimensionally black hole. This metric is horizonless (in
this sense this is a particle--like solution), without curvature
singularities and it has no signature change. The
solution~(\ref{SolucionFisica}) does have a conical singularity at
$r^{\prime}=0$ which can be removed by identifying the
$\phi^{\prime}$--coordinate with the period $T_{\phi^{\prime}}=2
\pi/(1+\tilde{q}^2_{m}/l^2)$~\cite{H-W}. It is well behaved, since
if $\tilde{q}_m$ approaches infinity, this period becomes zero,
while if $\tilde{q}_m$ approaches zero, this period goes to $2
\pi$, since the anti--de Sitter space has no angle deficit.
Finally, let us consider an alternative interpretation of this
magnetic solution. In the reference quoted above~\cite{Lemos}, the
authors have shown that the magnetic field source can be neither a
Nielson--Oleson vortex solution nor a Dirac monopole. Thus they
attempted to provide an interpretation of this magnetic solution.
Dias and Lemos interpreted the static magnetic field source as
being composed by a system of two symmetric and superposed
electric charges. One of the electric charges is at rest and the
other is spinning around it. In view of the symmetry of the
space--time this configuration is located at the origin of the
coordinate system.

We propose here another interpretation based on the similarities
of static Einstein--Maxwell theory for 2+1 dimensional
rotationally symmetric spacetimes and 3+1 dimensional axially
symmetric spacetimes~\cite{Cataldo1}. Let us refer to the static
magnetic fields in four dimensional general relativity.
Bonnor~\cite{Bonnor}, studying this topic, have considered axially
symmetric magnetostatic gravitational fields in empty spaces
generated from known electrostatic solutions. Bonnor noted here
that when we generate magnetostatic solutions from electrostatic
ones ``there is not an equivalence between sources of the static
electric and magnetic fields; by this is meant that whereas the
electrostatic field in empty space may be considered to arise from
point--charges, the magnetostatic field must arise from dipoles,
or from stationary electric currents"~\cite{Bonnor}. The above
remark may have profound implications for the nature of the
studied 2+1 dimensional magnetic spacetime. Effectively, Bonnor
generates a magnetostatic solution from a set of electrostatic
ones, for electric fields containing no matter or charge except at
singularities (see Eqs. (3.4) and (3.5) of the
Ref.~\cite{Bonnor}). This solution has two constants of
integration, representing the mass and the electric field
strength. The generated 3+1 dimensional magnetostatic solution has
physical sense only if we take zero the parameter representing the
mass; then the solution is regarded as referring simply to a
uniform magnetic field produced by a solenoid without
mass~\cite{Bonnor}. The similarity that happen between 2+1 and 3+1
dimensions is clear: the three--dimensional magnetic solution may
be generated from the electrostatic Ba\~nados--Teitelboim--Zanelli
black hole with the help of a duality
mapping~\cite{Cataldo1,Cataldo}. In this case the electric field
arises from a charged point mass (excluding interior solutions
from consideration). As we have shown the three--dimensional
magnetostatic gravitational field is really a one--parameter
solution, where the free parameter is only the integration
constant related to the magnetic field strength. Thus, the source
of the magnetic field may be considered a two dimensional
solenoid, i.e. a circular current. We prefer to locate this
current at spatial infinity, since the curvature is regular
everywhere.

We should note that the Bonnor solution with an uniform magnetic
field is valid for the case in which the cosmological constant is
vanished~\cite{Bonnor}. In our case the magnetic field is given by
\begin{eqnarray}
B(r) \sim \frac{1}{\sqrt{\frac{r^2}{l^2}+1}},
\end{eqnarray}
and is regular everywhere. From this expression we see that the
magnetic field at the origin has a maximum value, and at infinity
approaches to zero. This magnetic field is not a constant since
the cosmological constant is negative and then it acts as an
attractive gravitational force. This implies that the magnetic
lines held together near the origin.

{\it Note added.} In a recently appeared work the thin shell
collapse, leading to the formation of charged rotating black holes
in 2+1 dimensions, is considered~\cite{Olea}. In this context,
from physical considerations, the author singles out from the
solution~(\ref{H-W}) the case $M=-1$, since for this choice of the
parameter $M$ the magnetic solution does not exhibit a
pathological behavior. In this case a charged rotating thin shell
is interpreted as the analog to a solenoid carrying a steady
current, and then inside the thin shell the three dimensional
$M=-1$ magnetic static solution is valid, and the magnetic field
just vanishes outside the rotating thin shell.

\section{acknowledgements}
This work was partially supported by CONICYT FONDECYT N$^0$
1010485 (MC, SdC and PS), N$^0$ 1030469 (SdC and MC) and by
Ministerio de Educaci\'on through MECESUP grant FSM 9901 (JC). It
also was supported by the Direcci\'on de Investigaci\'on de la
Universidad del B\'\i o--B\'\i o (MC) and by grants 123.764-2003
of Vicerrector\'\i a de Investigaci\'on y Estudios Avanzados of
Pontificia Universidad Cat\'olica de Valpara\'{\i}so (SdC) and
UdC/DI 202.011.031-1.0 (PS and MC).

\end{document}